\documentclass[doublecol]{epl2} %for 2 columns style without line numbers
\usepackage{amssymb}
\usepackage{graphicx}
\usepackage{amsfonts}

\title{Non-exponential and oscillatory decays in quantum mechanics}
\author{Murray Peshkin\inst{1} \and Alexander Volya\inst{2} \and Vladimir Zelevinsky\inst{3}}
\shortauthor{M. Peshkin, A. Volya and V. Zelevinsky}

\institute{
\inst{1} Argonne National Laboratory, Physics Division, Argonne, Illinois 60439. \\
\inst{2} Florida State University, Physics Department, Tallahassee, Florida 32306. \\
\inst{3} National Superconducting Cyclotron Laboratory and Department of Physics\\
\inst{} and Astronomy, Michigan State University, East Lansing, Michigan 48824.
}
\abstract{
The quantum-mechanical theory of the decay of unstable states is revisited.  We show that the decay is non-exponential both in the short-time and long-time limits using a more physical definition of the decay rate than the one usually used.  We report  results of numerical studies based on Winter's model that may elucidate qualitative features of exponential and non-exponential decay more generally.  The main exponential stage is related to the formation of a radiating state that maintains the shape of its wave function with exponentially diminishing normalization.  We discuss situations where the radioactive decay displays several exponents.  The transient stages between different regimes are typically accompanied by interference of various contributions and resulting oscillations in the  decay curve.  The decay curve can be fully oscillatory in a two-flavor generalization of Winter's model with some values of the parameters.  We consider the implications of that result for models of the oscillations reported by GSI.}

\begin{document}
\maketitle

\section{Introduction}
This paper presents our perspective on the quantum-mechanical theory of decaying states.  In part of it we make use of Winter's solvable model, for which we have made numerical calculations.  For brevity and to keep the perspective in focus, we frequently characterize the results of those calculations, presenting few details.  Needed details will be published elsewhere. 

The exponential decay of unstable states is one of the most pervasive and most studied phenomena in microscopic physics, yet its quantum-mechanical theory remains obscure in many ways. 

The classical decay rate $R_c(t)$ is proportional to the available population of the unstable state and therefore to the classical survival probability $S_c(t)$,
\begin{equation}
R_c(t)\equiv -\frac{dS_c}{dt}=\Gamma\, S_c(t)  \leadsto R_{c}(t)=\Gamma\,e^{-\Gamma t}.       \label{1}
\end{equation}
The constancy of $\Gamma$ in Eq. (\ref{1}) is responsible for exponential decay law.
%\begin{equation}
%R_{c}(t)=\Gamma\,e^{-\Gamma t}.                \label{2}
%\end{equation}
This description is based on the assumption that only the population, not the structure, of the parent state changes over time while the mechanism of decay of every member of the population is constant.  More generally, the parent state could have substates, each with its own initial population and its own $\Gamma.$  Then  $R_c(t)$ would be a sum of decreasing exponentials with positive coefficients and it would decrease monotonically as function of  time. 
Experiments have observed exponential decay over many half-lives. A  detailed statistical analysis~\cite{silverman} shows that the exponential stage of beta-decay is indeed a random process.

No quantum-mechanical counterpart of Eqs.~(\ref{1}) with constant  $\Gamma$ is valid at all times.  
Many authors \cite{merzbacher98,Khalfin:1957,Khalfin:1958,schwinger60,newton61,terentev72,greenberg72,nicolaides77,peres80,garcia96,alzetta66,Baz:1969,ekstein71,Fonda:1978} 
have emphasized the generally non-exponential nature of decay laws  in quantum mechanics, along with the problems of experimental observation of deviations from a pure exponential decay. The discussion has recently been  revitalized by the reported 
%experimental 
observation of  decay oscillations during times comparable with the half-life in neutrino emission~\cite{GSI,GSInew}. That observation remains to be fully confirmed but it has 
%nevertheless 
stimulated active theoretical discussions, for example~\cite{Ivanov:2009, flam10, Ivanov:2010, pavlichenkov10, gal:2010}, with the contributors taking diametrically opposite views. 
Evidence for non-exponential decay in the case of $^{14}$C used for radioactive dating
has been claimed and debated \cite{aston12,nicolaides13,aston13}.

Textbook treatments of the subject vary 
%in their approach to approximate solutions of the Schr\"{o}dinger equation, 
but they appear all to be described accurately by the words of E. Merzbacher \cite{merzbacher98}:
%\begin{quote}
"{\sl the exponential decay law ... is not a rigorous consequence of quantum mechanics but the result of somewhat delicate approximations.}"
%\end{quote}
There is evidently something right about those delicate approximations because they do yield the correct exponential decay rates in some atomic physics cases where everything needed for the calculation is known.  However they are silent about the range of times during which the decay is exponential.  Moreover, they define the survival probability of the initial state $\Psi(0)$as
\begin{equation}
S(t)=\left |\langle \Psi(0)|\Psi(t) \rangle \right |^2.     \label{3}
\end{equation}
That definition
seems to us
unrepresentative of what experiments really measure.  $S(t)$  should be the probability of finding a parent nucleus or an excited atom at time $t,$ not necessarily the probability of finding the system in exactly the state represented by $\Psi(0).$

The presence of three regimes -- initial, exponential, and long-time inverse power law -- appears  to be a universal feature of the decay process. The transitions from one regime to another are accompanied by the interference of corresponding quantum amplitudes seen as oscillations on the decay curve. Below we review different decay regimes.  We also consider the possibility of more unusual oscillatory modes caused by quantum dynamics of populations of different substates.

\section{Winter's model}
The physics of decay
% in quantum mechanics
was clarified by Winter \cite{Winter:1961} with
%who introduced
a
%solvable
potential model that has features resembling those of a real physical system.
This model became a
useful tool to study non-exponential features in decay \cite{dicus02,QP}.
In Winter's model a particle of mass $m$ moves in one dimension between an impenetrable wall at $x=-1$  and $x\rightarrow\infty$ 
under the influence of  a confining
potential $V(x)=G\delta(x).$ 
%At time $t=0$, the wave function $\Psi(x,t)$ is confined to the interior, or parent, region $x\leq 0$.
Taking $\hbar=2m=1$ Winter's Hamiltonian and its energy eigenfunctions $|k\rangle$ 
are given by 
\begin{equation}
H_W=-\frac{\partial^2}{\partial x^2} + G \delta(x),
\end{equation}
\begin{equation}
\langle x|k\rangle= \sqrt{\frac{2}{\pi}} \left \{
\begin{array}{cc}
 \frac{\sin(\phi_k)}{\sin(k)} \,\sin\left [k(x+1)\right], & -1\le x<0 \cr
 \sin(kx+\phi_k), & x\ge 0
\end{array}
\right ..                                                  \label{25}
\end{equation}
where $E$ is the energy,  $k=\sqrt{E}$ is the asymptotic momentum,  states $|k\rangle$ are normalized as $\langle k|k'\rangle=\delta(k-k'),$  and  $\cot(\phi_k)=\cot(k)+{G/k}. $

At time $t=0,$ the wave function $\Psi(x,t)$ is confined to the interior, or parent, region $x \le 0.$  The survival probability is defined as
\begin{equation}
S_W(t)=\int_{-1}^{0} \left | \Psi(x,t) \right |^2\,dx.
\end{equation}
Winter solved the time-dependent Schr\"odinger equation for physically motivated values of the parameters by expanding the wave function in the complete set $|k\rangle $ to find that the decay rate $R(t)$ rises from zero at $t= 0$ in a time significantly shorter than the halflife, then settles into an exponential regime for many halflives, and finally goes over into the expected inverse power of $t.$  (See also more accurate calculations by Dicus et. al. \cite{dicus02} and in the textbook \cite{QP} based on the same model.)   

Winter's model presents an opportunity to test a consequence of Merzbacher's ``delicate assumptions'' Ref.~\cite{merzbacher98} that lead to the expansion of the wave function in poles at complex energies.  In our calculations we generally used that expansion, as did Winter.  However we also 
%solved the Schroedinger equation by 
applied direct numerical integration 
%in some cases 
and obtained identical results.  

%\section{Features of decay in quantum mechanics}

\section{The radiating state}
In our numerical calculations based on Winter's model, we have found that the parent 
%state 
wave function, i.e. the part of $\Psi(x,t)$ in the region $x < 0,$ rapidly approaches what we call the ``radiating state", 
\begin{equation}
\Psi(x,t)\simeq\Psi_R(x)\,e^{-\Gamma  t/2} e^{-i E_R t},                   \label{11}
\end{equation}
and remains there throughout the exponential stage.  The radiating state wave function $\Psi_R(x)$ is independent of the initial state
%wave function 
and it remains constant in shape during the exponential regime.   $E_{R}$ in Eq.(\ref{11}) is the average kinetic energy in $x < 0$ during the period of decay with rate  $\Gamma.$

\begin{figure}
\includegraphics[width=0.9\linewidth]{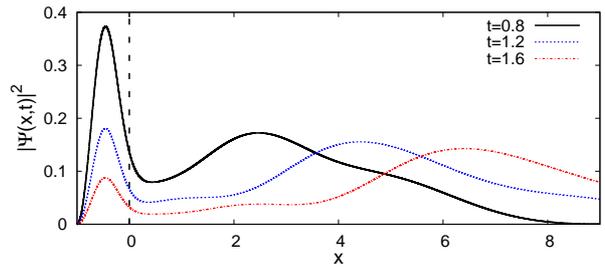}\caption{Probability distribution $|\Psi(x,t)|^2$
%Wave function of 
for a decaying state 
%shown as time snapshots 
at $t=0.8$, 1.2, and 1.6.
% Upper panel shows the
 Here $G=6.$\label{fig:Tsnapshots}}
%GG6/GG6_RT
\end{figure}

The radiating state provides the quantum-mechanical counterpart of the classical survival probability $S_c$ of Eqs. (\ref{1}) and it 
%appears to 
gives some  justification for a key assumption in quantum-mechanical treatments 
%which assume 
that the rate of change of the amplitude for being in the parent state is proportional to that amplitude.  Fig. \ref{fig:Tsnapshots} illustrates a typical example of a calculation. 
%based on Winter's model.  
The wave function for $x < 0$ retains its shape as function of $x$. 
%at several times.

\section{The earliest times}
That the decay rate at time $t=0$ must vanish has been proved under various assumptions \cite{Fonda:1978}, defining the survival probability as 
in Eq. (\ref{3}), which as argued earlier is generally unphysical.

The proper definition depends 
%in principle 
on the experiment used to detect the parent or daughter state but in practice that appears never to be an issue.  However that may be, let $P$ be the projector on states in which a parent atom 
or nucleus is present and no daughter is present.  The appropriate survival probability $S$ and decay rate $R$ are given by 
\begin{equation}
S(t)=\langle P \Psi(t) | P \Psi(t) \rangle,                       \label{6}
%\label{eq::a3}
\end{equation}
\begin{equation}
R(t)=-dS(t)/dt=
-2\, {\rm Im} \left \{ \langle P \Psi(t) H| (1-P) \Psi(t) \rangle \right \}.   
\label{der}
\end{equation}
If it is assumed that  $(1-P) \Psi(0)$ vanishes, then 
\begin{equation}
R(0)=0.
\label{10}
\end{equation}
This result is an exact consequence of the Schr\"odinger equation under the assumption, valid in the Winter model, that there is a time $t=0$ when the parent state is surely present and the daughter absent.   In real physics that assumption, which is explicit or implicit in all derivations of Eq.~(\ref{10}) of which we know, can never be exact and its applicability varies with  the application.  An unstable nucleus may be  created either by a reaction in some other channel whose participants are long gone at times of interest and need not be included in the calculation, or filtered by some experiment.  In either case the process takes a finite time but that time is typically very short compared to times of physical interest and Eq.~(\ref{10}) is an excellent approximation. Then $R(t)$ rises from 0 at $t=0$ before settling into an exponential or 
%possibly 
a sum of a few exponentials.  For a long-lived resonance in a scattering process, Eq.~(\ref{10}) lacks even approximate validity.  
%We note that 
This derivation of Eq.~(\ref{10}) agrees with earlier work that 
%defined survival through 
used Eq.~(\ref{3})
%.  To see that it is necessary only to define 
if $P$ is defined as the projection on  $\Psi(0).$

\section{Effective non-Hermitian Hamiltonian\label{sec:CSM}}
Starting from diverse approaches
\cite{Fonda:1978,muga96},
the time behavior has been  expressed as a squared Fourier integral $S(t)=|{\cal F}(t)|^{2}$, of some energy-dependent amplitude,
\begin{equation}
{\cal F}(t)=\int dE\,\exp(-iEt)D(E).                                                  \label{15}
\end{equation}
%The function $D(E)$, closely related to the S-matrix,
%being continued to the complex energy plane.
%
%The time behavior is typically  expressed as a squared Fourier integral $S(t)=|{\cal F}(t)|^{2}$, of some energy-dependent amplitude,
%\begin{equation}
%{\cal F}(t)=\int dE\,\exp(-iEt)D(E).                                                  \label{15}
%\end{equation}
The function $D(E)$, closely related to the S-matrix,
%being continued to the complex energy plane,
has resonance poles,
${\cal E}_{r}=E_{r} -\,\frac{i}{2}\,\Gamma_{r},$                                 
% \label{16}
%$\end{equation}
in the lower part of the complex energy plane, $\Gamma_{r}>0$.

The exponential time evolution
of an initial quasi-stationary state corresponds to a single complex pole and 
%leads 
to
%\begin{equation}
%\Psi(t)=\,\exp\left[-\,\frac{i}{\hbar}\,\left(E_{0}-i\,\frac{\Gamma}{2}\right)t\right]\Psi(0), \label{10}
%\end{equation}
the Lorentzian  energy spectrum of the decaying state,
%given by the Lorentzian,
\begin{equation}
|\Psi_{E}|^{2}\propto \,\frac{\Gamma/2}{(E-E_{r})^{2}+\Gamma_r^{2}/4},               \label{12}
\end{equation}
Many features of time-dependent decay, including the short-time evolution, interference between resonances and decay at remote times, can be explored
%and understood
using the description in terms of the {effective non-Hermitian Hamiltonian}.
This approach based on the Feshbach projection \cite{feshbach} has found wide applicability in various branches of science; examples and references can be found in a review article \cite{AZ11}. Explicit dynamics
%of time evolution
of many-body states in realistic nuclear systems was studied
 by this method in Ref. \cite{Volya:2009}.

In this approach 
the Hilbert space is separated into intrinsic (or parent) part  $P$ and an external part, 
the states with the asymptotics of continuum {channels}. 
%In this approach t
%The total Hilbert space is separated into intrinsic states and external part,
%(Feshbach projector $P$ \cite{feshbach}) and external part, $1-P$,
%the states with the asymptotics of continuum $-$ {\sl channels}. 
Then
%In this approach
the function $D(E)$ in Eq.~(\ref{15}) emerges from the effective propagator that describes the evolution  in the  subspace $P$,
\begin{equation}
{\cal G}(E)=\,\frac{1}{E-{\cal H}}, \quad {\cal H}=H-\,\frac{i}{2}\,W.                                       \label{17}
\end{equation}
%The Feshbach projection \cite{feshbach} is used to account for the dynamical components of the full space, thus retaining the propagator (\ref{17}) exact. As a result of this projection, the
The  effective Hamiltonian ${\cal H}$  in Eq. (\ref{17}) is energy dependent and, for energy $E$ above thresholds, non-Hermitian, where
%\begin{equation}
%{\cal H}=H-\,\frac{i}{2}\,W.                                             \label{18}
%\end{equation}
the anti-Hermitian part $W$ describes
%reflects the decay associated with
the loss of flux from the intrinsic space. The
%conservation of probability and
unitarity of the scattering matrix requires $W$ to be factorized,
\begin{equation}
\langle 1 | W |2 \rangle=\sum_{c\,({\rm open})}A^{c}_{1}A^{c}_{2}.                          \label{19}
\end{equation}
The amplitudes $A^{c}_{1}\equiv \langle c, E | H |1\rangle$
%, which can be taken real for the system with time-reversal invariance,
are the matrix elements of the original Hamiltonian between an intrinsic state $|1\rangle$ and the channel state $|c, E\rangle.$ The channel state is labeled here by the asymptotic energy $E$ and all additional quantum numbers are combined in label $c.$  For convenience, in this formalism the channel states are normalized by delta function of energy $\langle c, E | c', E'\rangle=2\pi\, \delta_{cc'}\,\delta(E-E').$ 
Therefore, the kinematic factors (density of states in the
continuum for a given channel) are included in these amplitudes so that they depend on running energy $E$
and vanish at the threshold of a given channel; only channels which are open at a given energy
contribute to this on-shell part of the Hamiltonian.
The resonances emerging from the poles of ${\cal G}(z)$ in the complex energy plane determine the analytic structure of the function $D$ in Eq. (\ref{15}).

\begin{figure}
%\onefigure{csmfft}
\includegraphics[width=0.9\linewidth]{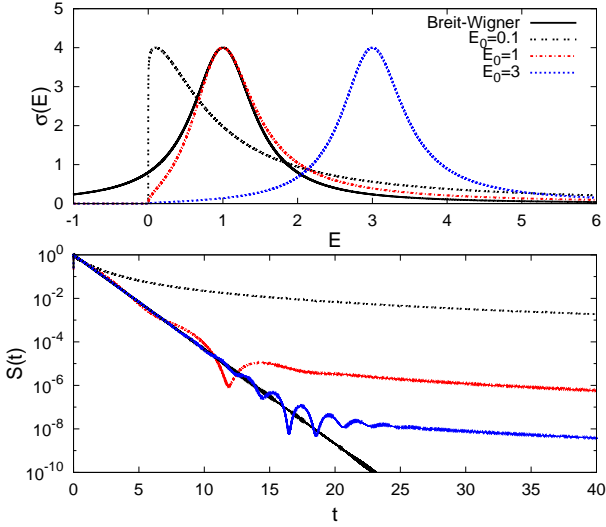}
\caption{Single resonance near threshold. Top  {panel}
% shows the 
energy-dependent scattering cross section off a resonance. Lower  {panel}
% shows 
the
% corresponding 
survival probability as a function of time. The Breit-Wigner resonance with the %characteristic
Lorentzian shape, Eq. (\ref{12}), would correspond to pure exponential decay. 
%at all times. 
The realistic resonances
at different positions $E_0$
%are also shown, these resonances
are modeled with energy-dependent Breit-Wigner $s$-wave kinematics and
%which assures that the
cross section going to zero at threshold.
%energy $E_{\rm th}=0;$ see discussion in Sec.~\ref{sec:CSM}.
\label{fig:csmfft}}
\end{figure}

Let us illustrate
%the time-dependence of the decay using
this approach with an example
%simple situation
of one intrinsic state coupled to the continuum, Fig.~\ref{fig:csmfft}. The survival amplitude for this
state is given by the expectation value of the propagator (\ref{17}) in the initial state $\Psi(0)$ that is assumed to be in the intrinsic space,
\begin{equation}
{\cal G}(E)=\frac{1}{E-E_0+\frac{i}{2} \Gamma (E)},\quad {\rm where} \quad \Gamma(E)=A^2(E). \label{20}
%\label{eq:EBW}
\end{equation}
%represents Fermi Golden rule.  
The energy dependence of amplitudes at low energies follows from decay width being proportional to the density of states; for spherically symmetric s-wave decay
$
\Gamma(E)\propto \int d^3 k \,\delta(E-k^2)\propto \sqrt{E}.
$
%in the initial state $\Psi(0)$ that is assumed to be in the intrinsic space,
%\begin{equation}
%Then $D(E)=(1/2\pi){\cal G}(E).$ 
%\langle \Psi(0)|{\cal G}(E)|\Psi(0)\rangle.  \label{21}
%\label{eq:D}
%\end{equation}
If the state is far from the threshold and the energy dependence of $\Gamma$ is ignored, this propagator ${\cal G}(E) =2\pi D(E)$ represents the exponential time evolution of the initial state with a
%characteristic
Breit-Wigner 
%shape of the 
cross section, Fig.~\ref{fig:csmfft}.
%as is seen below.
%As a result of this non-exponential features in the decay emerge, we discuss them in what follows.

\section{Pre-exponential dynamics}
At short times the wave function can contain multiple components reflecting the structure
%of the initial state being
distributed among many resonant poles. The pre-exponential dynamics involves transitions between these states and radiation with different exponential rates finally
%that takes place until the majority of the exponential terms decay
leaving a single exponential term corresponding to the pole closest to the real axis.
%These two features are distinct as i
Internal transitions
%differentiated by measured
influence the survival probability 
%according to the definition 
of Eq. (\ref{3})
whereas the total radiation is measured by Eq. (\ref{6}).

\begin{figure}[h]
\includegraphics[width=0.9\linewidth]{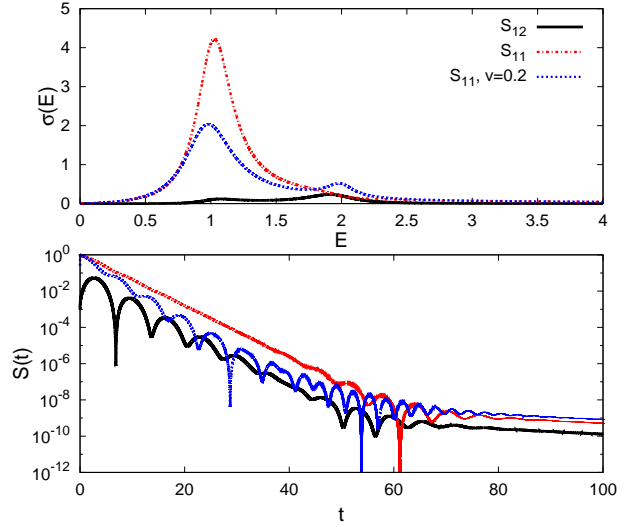}
\caption{Two-state model 
%defined by 
for the effective Hamiltonian (\ref{23}). 
%As in Fig. \ref{fig:csmfft}, t
Top panel: 
shows the 
energy-dependent cross section.
% of a resonance. 
Lower panel: 
%shows 
the survival probability as a function of time.
%For the model it is assumed that
Here $\epsilon_1=1$, $\epsilon_2=2;$ $\Gamma_1(E)=\Gamma_2(E)=\sqrt{0.1E}$.
%, is taken for both models.
The curve 
%labeled as 
$S_{11}$ shows the survival probability for the first intrinsic state, and $S_{12}$ 
%shows 
the probability for transition $2\rightarrow 1$; $v=0$ in these cases. The third curve illustrates the behavior $S_{11}$ in the mixed case $v=0.2$ for amplitudes having opposite signs.
\label{fig:csmfft2}}
\end{figure}

Consider 
%a system of 
two overlapping resonances described by the general non-Hermitian Hamiltonian,
%given by the matrix,
%see Ref. 
\cite{VZ03},
\begin{equation}
{\cal H}=\left(\begin{array}{cc}
\epsilon_{1}-(i/2)\Gamma_{1} & v-(i/2)A_{1}A_{2} \\
v-(i/2)A_{1}A_{2}    & \epsilon_{2}-(i/2)\Gamma_{2} \end{array}\right).    \label{23}
\end{equation}
Here $\epsilon_{1,2}$ are diagonal elements of the Hermitian part $H$ and $v$ is the internal mixing; the non-Hermitian part $W$ contains
$\Gamma_{1,2}=A_{1,2}^{2} $
where the amplitudes $A_{1,2}$ are real and in general dependent on running energy $E$.
The cross section of the resonance reaction and
%time dependence of
the survival probability for this model are shown in Fig.~\ref{fig:csmfft2}.
%This system of decaying resonances that interact both through real and imaginary components has a number of remarkable features that are also generic for the physics of unstable systems, see details in Ref. \cite{VZ03}.

In the limit of separated resonances far from the threshold, the behavior and decay curves are nearly identical to those for a single state, Fig.~\ref{fig:csmfft}.
When the states are close to each other, their mixing and interference occur via both Hermitian and non-Hermitian part resulting in the oscillatory modulation of the survival probability with beat times $ t\sim 1/|\epsilon_1-\epsilon_2|.$ Such oscillations reflect internal transitions in the decaying system with two radiating states mixed by internal interactions, 
%as is seen in 
Fig.~\ref{fig:csmfft2}.
%This pattern between the two internal states is generally not associated with decay from space $P$; o
Oscillations in the decay rate can emerge only if the two radiating states have different decay rates.
%In the limit $\Gamma\gg\Delta$, we come to the super-radiance phenomenon, when
%one of the states accumulates the whole width $\Gamma$ while the second one becomes a {\sl bound state
%embedded into continuum} \cite{friedrich85,VZ03}. In this limit the states effectively decouple and behave as those discussed in Fig.~\ref{fig:csmfft}, with the exception that the super-radiant state is likely to be distorted by the presence of the threshold and thus may exhibit no exponential decay at all.
At remote times determined by the decay width and the distance to the threshold, the model recovers
the asymptotic power-law behavior discussed below.

%%%%%%%%%%%%%%%%%%%%%%%%%%%%%%%%%%%%%%%%%%%%%%%%
In Winter's model
%the overlap between the intrinsic
%states and the continuum eigenstates is given analytically by
%\begin{equation}
%\langle n|k\rangle=\frac{2n\sqrt{\pi}}{k^{2}-n^{2}\pi^{2}}\frac{k\sin(k)}{\sqrt{k^{2}+k\,
%G\sin(2k)+G^{2}\sin^{2}(k)}}.                                  \label{26}
%%\label{eq:overlap}
%\end{equation}
the internal dynamics is determined by matrix elements of the evolution operator,
\begin{equation}
M_{nn'}(t)=\langle n|e^{-iHt}|n'\rangle=
\int_{0}^{\infty}e^{-ik^{2}t}\langle n|k\rangle\langle k|n'\rangle dk,  \label{27}
%\label{eq:A}
\end{equation}
where a complete set of intrinsic states for $x\in[-1,0]$ is defined as  
%and remains in one of the eigenstates of this potential
$ \langle x| n\rangle=\sqrt{2}\sin[n\pi (x+1)]$ with $n=1,2,\dots .$
The survival probabilities for individual states are  $S_{nn}(t)=\left|M_{nn}(t)\right|^{2}.$
The analytic properties of the functions $D_{nn}=\langle n|k\rangle\langle k|n\rangle$ are seen from Eq. (\ref{25}).
The poles $E_r=k^2_{r}$ correspond to the roots of the equation $k^{2}+k\, G\sin(2k)+G^{2}\sin^{2}(k)=0.$
%The integral in Eq. (\ref{27}) cannot be closed with a semicircle
%in the complex plane because it diverges in the first and third quadrants.
%However, the pole (resonant) contribution can be separated by considering a contour
%$OABO$ as shown in Fig. \ref{fig:Integration-contour}. Here, a path
%$OA$ from origin $O$ to a remote point $A$ represents the original
%integral in Eq. (\ref{27}). The integral along a remote 45$^{\circ}$
%arc is zero for $t>0$ due to exponential suppression. Finally, integrating
%along the diagonal path $BO$ where ${\rm Re}(k)={\rm Im}(k)$ leads to a Gaussian
%integral. Here the exponential term is real.
%\begin{figure}
%\includegraphics[width=0.9\linewidth]{contour}\caption{Integration contour\label{fig:Integration-contour}}
%\end{figure}
In Eq. (\ref{15}) the integration contour in the complex energy plane for $t>0$ should be drawn in the fourth quadrant
and closed on the real axis at the threshold point $E=E_{{\rm th}}$.
%of the lower boundary of the energy spectrum.
Integration in Eq. (\ref{27}) (where $E_{{\rm th}}=0$)
%conducted this way using the residue theorem
gives rise to two types of contributions,
%in the amplitude:
the terms responsible for exponential decay from the
poles $k_r$ and the non-exponential component $M_{nn'}^{(NR)}(t)$   from the Gaussian-type
integral along the vertical boundary of the fourth quadrant (${\rm Re}\,E=0$),
%Thus,
\begin{equation}
M_{nn'}(t)=\sum_{k_{r}}M_{nn'}^{(r)}\, e^{-ik_{r}^{2}t}+M_{nn'}^{(NR)}(t).  \label{28}
%\label{eq:AC}
\end{equation}
Here the resonance amplitudes, 
%(those of exponential terms) ,
%\begin{equation}
$M_{nn'}^{(r)}(t)=-2\pi i\,{\rm Res}(k_{r})$,  given by the residues at the poles,
%\label{29}
%\end{equation}
%and position of poles
 determine exponential dynamics, $\Gamma_{r}\equiv1/\tau_{r}=-2{\rm Im}(k_{r}^{2}).$
The non-resonant contribution follows from the direct integration; in the large time limit 
%we obtain
\begin{equation}
M^{(NR)}_{nn'}(t)=\,\frac{1+i}{\pi^{5/2}\sqrt{2}}\,(1+G)^{2}\,\frac{1}{nn'}\,\frac{1}{t^{3/2}}. \label{30}
%\label{ANR}
\end{equation}

%The short-time survival probability $S_{11}(t)=\left|M_{11}(t)\right|^{2}$
%of the lowest intrinsic state $n=1$ is shown in Fig. \ref{fig:Short-time-survival}.
%As said, initially the decay amplitude contains many terms which lead to non-exponential decay. In order to highlight
%the role of these terms we compare the survival probability with the pure exponential
%decay given by the first pole $\tau_{1}$ and with the result that
%includes the first two poles $\left|M_{11}^{(1)}e^{-ik_{1}^{2}t}+M_{11}^{(2)}e^{-ik_{2}^{2}t}\right|^{2}.$
%The interference between the poles results in the oscillatory behavior at short times. Such effects can be responsible,
%for example, for high-frequency part of bremsstrahlung in the radiation accompanying alpha-decay \cite{bertulani99}.
%\begin{figure}
%\includegraphics[width=0.9\linewidth]{s6_1_bw_small}
%\caption{Short-time survival probability in Winter's model. The parameter $G=6$
%was selected. The main part of the decay curve is represented
%by the single pole term, $k_{1}=2.75794-i\,0.140433,$ $\tau_{1}=0.65$;
%the amplitude of this component is $M_{11}^{(1)}=0.99068-0.06804i$,
%which at $t=0$ provides 98.6\% of normalization. The contribution from
%the second pole, $k_{2}=5.71348-i\,0.370148$, or $\tau_{2}=0.12$) with
%$M_{11}^{(2)}=0.00806+i0.03016$ is small but clearly influences the
%decay at short times. \label{fig:Short-time-survival}}
%\end{figure}

\begin{figure}
\includegraphics[width=0.9\linewidth]{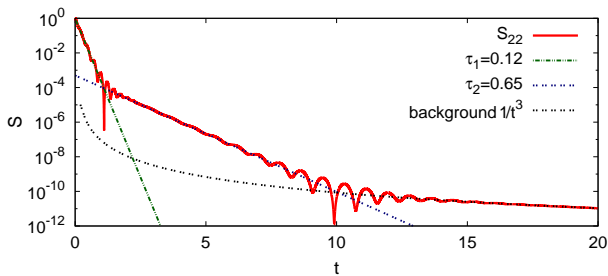}\caption{Survival probability $P_{22}(t)$ as a function of time. For
%the value of
%the delta strength 
$G=6$,
the two lowest complex poles are at $k_{1}=2.75794-i\,0.140433$ ($\tau_{1}=0.65$) and $k_{2}=5.71348-i\,0.370148$ ($\tau_{2}=0.12$), their resonant contributions are shown along with the non-resonant background from Eq. (\ref{30}).
\label{fig:P2}}
\end{figure}

During the initial evolution
%the behavior of
the survival probability is
%represented as
a collection of radiating exponents, the sum over poles $k_r$ in Eq. (\ref{28}). This is highlighted in Fig.~\ref{fig:P2} where the survival probability is shown for the initial state $n=2$. This state has only a small $r=1$ component $M^{(1)}_{22}=0.0005$ and a very large $r=2$ component $M^{(2)}_{22}=0.9917.$
Thus, at short times the decay follows $\exp(-t/\tau_2)$ until the resonant component corresponding to $\tau_2$ dies out. The amplitudes associated with  $r=1$ and $r=2$ poles become similar
%, which happens here
at $t \approx 1$ in Fig.~\ref{fig:P2}; then the two decay modes interfere. The transition to the pure $\exp(-t/\tau_1)$ decay law ends the pre-exponential dynamics.

\begin{figure}
\includegraphics[width=0.9\linewidth]{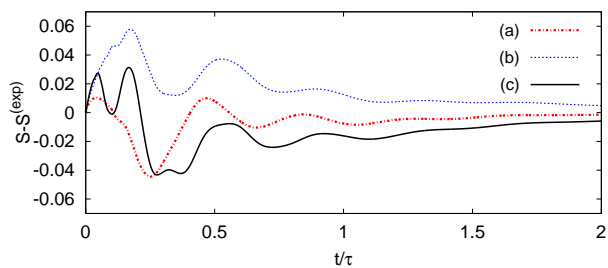}
\caption{
%Deviation from exponential decay is shown as a d
Difference between the actual survival probability and
%the exponential
$S(\rm{exp})=\exp(-t/\tau),$ where $\tau=\tau_1$ is the leading term in Eq. (\ref{27}),  $G=6.$ The time is expressed in units of the lifetime $\tau.$  Curve (a): 
%shows the 
deviation of  $S_{11}(t)$ from the exponent. Curve (b) corresponds to $S(t)$ from Eq. (\ref{6}) where intrinsic space $P$ is defined as $x\in[-1,0]$. Curve (c) is the same as (b) but for the initial wave function with finite $\Delta E$, $\Psi(x,0)=\sqrt{8/3}\, \sin^2{(\pi x)} .$  \label{fig:P1_D}}
%winter/winter_fft/s6_1_os.gnu
\end{figure}

The 
%type and the duration 
details of the pre-exponential regime depend on the disposition of poles, on the exact form of the initial wave function, as well as on the quantity observed as a decay signal. The 
%deviation from the exponential behavior %This is illustrated , where
resulting behavior is shown in Fig. \ref{fig:P1_D}.

\section{Decay rate at long times\label{power}}
Deviations from exponential decay are also expected in the post-exponential long-time limit, when
%For example in Winter's model in Eq.~(\ref{28})
%at long times
% all 
resonant sources dry out leaving only the non-resonant component with its characteristic power-law time dependence 
 %survival probability.
% This can be 
as seen in Fig.~\ref{fig:P2}.
% where
%the longest-lived resonant component associated with $\tau_1$ diminishes exponentially and eventually the non-resonant contribution with the power-law decay, Eq.~(\ref{30}), takes over.
 The transitional region around $t=10$ is associated with the interference of exponential and power-law terms.

The non-exponential long-time behavior comes from the fact that 
%for a realistic finite-sized system %(atom, molecule, nucleus, particle, atoms in traps, ...)
the energy spectrum has a lower bound.
The exponential decay implies Lorentzian in 
energy spectrum
%Eq. (\ref{12})
%that reveals 
with an unphysical tail to infinitely low energies.
As the physical spectrum of a given decay channel
starts at a threshold, let say $E_{{\rm th}}=0$, the exponential stage with 
%the lifetime 
$\tau \sim 1/\Gamma$
is expected to end at $t\sim t_{L}$ when
\begin{equation}
e^{-t_{L}/\tau}\sim \,\frac{E_{0}}{\Gamma} \quad \leadsto \quad t_{L}\sim \tau\,\ln\left(\frac{E_{0}}{\Gamma}\right).
                                                          \label{13}
 \end{equation}
%From mathematical perspective at
At $t \rightarrow \infty$, the behavior
of $D(E)$ at threshold, $E\rightarrow E_{{\rm th}}$, 
related to  the physics of non-resonant particles near the  lowest energy, defines the time dependence of decay.
%This behavior is determined by the physics of non-resonant particles at very low energy.
%with well-known universal properties.
%Asymptotic of
Close to threshold, $D(E)$  is determined  by the density of states in the continuum,
% therefore
typically $D(E)\propto (E-E_{{\rm th}})^{\lambda}.$
%In the case
For  a neutral particle with orbital momentum $\ell$
leaving a spherically symmetric potential, $\lambda=\ell+(1/2)$.
%, where $\ell$ is the orbital momentum of the
%outgoing particle. 
This contribution to ${\cal F}(t)$ is proportional to $1/t^{\lambda+1}$, so that
the asymptotic power law for $|{\cal F}(t)|^{2}$ is $\propto 1/t^{2\lambda+2}$, or $1/t^{3}$ for the $s$-wave.

%Study in
Fig.~\ref{fig:csmfft}  illustrates how the threshold at $E_{\rm th}=0$ distorts the Lorentzian
% tail 
and eliminates exponential decay.
%at remote time.
%The study is done assuming
Here the $s$-wave behavior of the decay width, $\Gamma(E)\propto \sqrt{E}$, is assumed. The transition time $t_L$
%to the power law behavior
 depends on the distance to the threshold as illustrated by Fig.~\ref{fig:csmfft}. 
 %several curves illustrate the decay curve of a resonance placed at different distances
 %away
% from the threshold.
The decay width $\Gamma(E)$ is proportional to the phase space volume and thus also scales as a power law. The power-law feature is
correctly recovered in the effective Hamiltonian approach leading to $S(t)\sim 1/t^{2\ell+3}.$

% in the long-time limit, due to oscillations of the exponent only the real part of Eq. (\ref{21}) contributes
%\begin{equation}
%D(E)\rightarrow \frac{\Gamma(E)}{(E-E_0)^2+\Gamma(E)^2/4} \sim (E-E_{\rm th})^{l+1/2}, \label{22}
%\end{equation}
%leading to $S(t)\sim t^{2l+3}.$

Alternatively, we can say that the initial internal state, being coupled 
%by interactions 
to the continuum, is not stationary and thus contains the low-energy components of the continuum; the power-law decay emerges due to the slowest particles present in this wave function \cite{newton61,torrontegui10}.
%The Feshbach projection method allows to recover a full-space wave function, where component $\langle E|\Psi(0)\rangle $ associated with eigenstate state $|E\rangle$ of energy $E$ is given by an operator $A/ (E-{\cal H})$ therefore
%$D(E)=\langle \Psi(0)|E\rangle \langle E| \Psi(0)\rangle \sim (E-E_{\rm th})^{l+1/2}.$
%The non-resonant component can be thought of as being due to a non-interacting free-space
%wave function for scattered states; this component decays with the time dependence
%following a power law.
The outgoing background can be noticed in Fig. \ref{fig:Tsnapshots}.
In this logic, see also \cite{newton61},
one can say that at time $t$ the number of free particles of mass
$m$ and velocity $v$ in the region $\Delta x \sim 1/{(mv)}$ is
\begin{equation}
S(t)\propto\frac{\Delta x}{vt}\sim \frac{1}{mv^{2}t}=\frac{2}{E_{0}t}\propto\left|\Psi(t)\right|^{2}.
                                                                                   \label{14}
\end{equation}
This semiclassical argument implies a $1/t$ power-law for the survival probability.
In quantum mechanics the features of low-energy scattering with $1/t^{3}$ for $s$-wave should be used
in place of such semiclassical arguments.
%given above.
%While 
Even for a very small non-resonant component in the initial state,
% can be very small, because of its power-law time dependence this component 
it decays slower than the 
%exponentially decaying 
resonant part. At time $t\sim t_{L}$ the resonant and non-resonant components become comparable
and for $t>t_L$ decay becomes non-exponential.
Because of a tiny amount of decaying material left after many lifetimes, this stage is hard to observe
experimentally. But in the transitional region the resonance and threshold contributions
are of the same order and interfere so that the transition to the power law is accompanied by characteristic
{oscillations} seen in numerical simulations.

\section{Flavor oscillation model}
On the wave of the GSI experiments \cite{GSI,GSInew} it has been speculated that mixing of two close-in-energy radiating states could lead to exponential decay modulated by oscillations.  
%One has to distinguish this situation from that in our discussion of internal dynamics.
%in Fig. \ref{fig:G6}. 
The main difference with  our previous  discussion
% of internal dynamics 
is the presence of two intermixed
{decay channels}.
%Here we examine this possibility.
We consider a
%$2\times 2$
flavor-mixing model
\begin{equation}
\left(\begin{array}{c}
\nu_{+}\\
\nu_{-}
\end{array}\right)=\left(\begin{array}{cc}
\cos\theta & \sin\theta\\
-\sin\theta & \cos\theta
\end{array}\right)\left(\begin{array}{c}
\nu_{1}\\
\nu_{2}
\end{array}\right),     \label{36}
%\label{eq:fb}
\end{equation}
where + and - subscripts denote upper and lower flavor states and  subscripts 1 and 2
%in Eq. (\ref{36})
denote mass eigenstates.
The kinetic part $\hat{K}$ of the Hamiltonian is diagonal in the mass eigenstates and can be written
%in a matrix form
as
%\begin{equation}
%\hat{K}=
%\left(
%\begin{array}{cc}
%-\frac{\hbar^2}{2m_{1}}\frac{d^{2}}{dx^{2}}  & 0 \\
%0 & -\frac{\hbar^2}{2m_{2}}\frac{d^{2}}{dx^{2}}
%\end{array}
%\right) + \Delta \sigma_z
%\label{38}
%\end{equation}
\begin{eqnarray}
\hat{K}\nu_{1,2}(x)=\left [ -\frac{\hbar^2}{2m_{1,2}}\frac{d^{2}}{dx^{2}}\pm\Delta \right ] \nu_{1,2} (x),
\label{38}
\end{eqnarray}
where $\Delta$ 
% represents 
is the difference in the rest masses.
We assume that the potential barrier is diagonal in the flavor basis,
\begin{equation}
\hat{V} \nu_{\pm}(x) =G_{\pm}\,\delta(x)\,\nu_{\pm}(x).   \label{37}
\end{equation}

For equal masses and identical potentials for both flavors,
%If the potential is the same and the masses are equal $m_1=m_2\equiv m$ then
the problem is diagonal in the mass eigenstates and oscillations occur only between the flavor basis states which is not related to the decay process.
%This emphasizes that w
While the flavor oscillations would be evident with definition (\ref{3}), there are no decay oscillations associated with the probability to find the particle, independently of flavor, in the intrinsic space. In order for non-trivial oscillations to occur with the realistic  definition (\ref{6}),  there should be a substantial difference between the decay rates into the two flavor states.
The largest non-exponential effect is expected at the maximum  mixing  angle $\theta=45^\circ.$ With $G_{+}$ fixed, we explore the extreme situations with $G_{-}=0$ and $G_{-}=\infty.$

\begin{figure}
\includegraphics[width=0.9\linewidth]{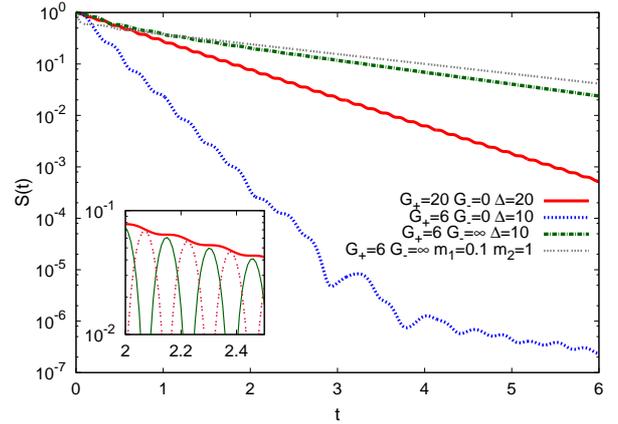}\caption{Total survival probability as a function of time for the two-flavor 
%neutrino 
model.  The mixing angle is 
%fixed at 
$\theta=45^\circ.$ Other parameters of the Hamiltonian 
%$\hat{H}=\hat{K}+\hat{V}$ in Eqs. (\ref{36}) and (\ref{38}) 
are indicated with labels. The initial state is assumed to be in the upper flavor 
%state 
with the $n=1$ wave function of an infinite box. Inset shows oscillations of the survival probability for two flavor components and the combined 
total for $G_+=6,\,G_-=\infty,\,\Delta=20$. 
%set of parameters.     
\label{fig:n_osc}}
%figs/n_osc.eps
\end{figure}

In Fig.~\ref{fig:n_osc} the
%survival 
probability to find a particle of any flavor in the $x<0$ region is shown as a function of time for various parameters.
All curves exhibit exponential decay. The average lifetime is governed by the combination of the barrier strengths.
For $G=6$ in Winter's model the mean lifetime $\tau=0.65$;  for the two-flavor model with equal masses and one flavor state not held by the barrier, $G_{-}=0$, the lifetime drops to $\tau\approx 0.25 $ ( {dotted line}); 
%in  Fig.~\ref{fig:n_osc}). Due to fast decay t
the post-exponential regime is quickly reached around $t=4$. 
%in this case. 
In the opposite limit with $G_{-}=\infty$, the forbidden decay in the lower flavor state extends the lifetime to $\tau\approx 2$ which is nearly the same for 
%the case of 
equal masses $m_1=m_2=0.5$ and $\Delta=10$ (dashed-dotted line) and for 
%the case of 
$m_1=0.1\,,\,\, m_2=1,$ and $\Delta=0$ ( {double-dotted line}).

%The exact choice of the initial wave function is important  for this limit, for our examples we assumed the wave function to be in the
%upper flavor state at $t=0$.

%In Fig. \ref{fig:n_oscx} we examine separately the case of $m_1=m_2=0.5$ and $\Delta=10.$ The initial oscillatory modulation of the exponential
%behavior is seen; this oscillation extends to about one mean lifetime. In addition to the total survival probability the plot contains the survival
%probability of each flavor component. Rapid and large scale oscillation of individual flavor components, the internal beat between quantum states, is
%the cause of oscillations in the total survival probability.

%It is remarkable that i
In the limit where the decay in one of the flavor states is not hindered by a potential, $G_{-}=0$, the flavor oscillations lead to the continuous 
%oscillatory 
modulation of the total (sum of both flavors) survival probability ( {solid line and dotted} line in Fig.~\ref{fig:n_osc}). Unlike in 
%all other 
previous situations, these oscillations extend over multiple lifetimes
%.  The modulating oscillations are 
being determined by the mass difference $\Delta$; 
%corresponding to 
the period $\pi/\Delta$ is seen 
%of the underlying flavor oscillations shown 
on inset.   
%In our view t
The presence of two decay channels is the principal difference in this case.

%\begin{figure}
%\includegraphics[width=0.9\linewidth]{figs/nxG6G100.eps}\caption{The total survival probability $S$ and the survival probabilities in the upper and
%lower flavor states as a function of time are shown for the two-flavor neutrino model.
%The parameters of the model are $m_1=m_2=0.5,$ $\Delta=10, $ $G_{+}=6, $ $G_{-}=\infty$ and $\theta=45^\circ.$ The initial state is assumed to
%be in the upper flavor state with the sine wave function for $n=1.$
%\label{fig:n_oscx}}
%\end{figure}

\section{Conclusions}
Although many elements of our discussion are well known, being spread over the literature, 
%we believe 
it is  useful to revisit
the 
%ubiquitous 
typical features of quantum decay of unstable states.
%The standard exponential decay, that covers many lifetimes is the main part of the time evolution.
%During this time the effective intrinsic wave function of the decaying state
%is stable forming what we called the {\sl radiating state} that has an exponentially decaying normalization.
%
In recent literature the nature and universality of the exponential decay has been questioned. 
Is exponential decay a
%process a 
result of some ``delicate" approximations that may 
%easily 
break down in certain limits? Is it possible to have an oscillatory decay behavior?
We have
examined several models, targeting their general analytic properties as well as showing exact numerical solutions.  
Our main conclusions can be summarized as follows:
%\begin{itemize}

$\bullet$ The formation of the radiating state that corresponds to the pole of the decay amplitude closest to the real axis is preceded by a short-time stage with low decay probability. 
The short-time limit can be sensitive to the details of the preparation of the unstable state (the experimental attempts were inconclusive \cite{norman88,norman95,scates03}).
However, the radiating state concept appears to be universal.

$\bullet$ The general  theoretical approach based on the effective  non-Hermitian Hamiltonian highlights the coupling of internal and external features.
%The internal dynamics of unstable complex systems is seamlessly coupled to decay channels, commonly leading to a single form of the dynamical evolution: exponential decay via radiating state. 
%%Thus, while the t
%The complicated time evolution of certain states 
%%undergoes a continuous complex behavior this behavior 
%has no observable effect upon the flux in outgoing channels. The theoretical approach based on the non-Hermitian effective Hamiltonian highlights the coupling of internal and external features.

$\bullet$ The energy dependence of the decay amplitudes, in particular due to the proximity
%of the branch points corresponding to
of  channel thresholds, is responsible for the power-law decay tail that in the limit
of long time wins over the main exponent
while the transitional stage 
%between the exponential regime and slower power-law decay
reveals 
%the second type of 
oscillations.

$\bullet$ The two-flavor model 
%with two open channels
shows that, for some sets of parameters, the entire decay curve is essentially oscillatory. Thus the possibility of observing oscillations in the decay process cannot be excluded. 
This result goes beyond the common scenario in which two states of the parent system are mixed by an external field so that their amplitudes in the wave function oscillate 
%in time 
and the decay curve may oscillate with them.  Here the two are mixed only indirectly by their coupling to the daughter states and the oscillations are not dependent upon external fields. 
However, a more realistic model will be needed to determine whether the analogous coupling
in neutrino emission causes similar oscillations.

\acknowledgements
This work is supported by the U.S. Department of Energy, Office of Nuclear Physics, under contracts DE-AC02-06CH11357 and DE-SC0009883, and by the NSF grants 
PHY-1068217 and PHY-1404442. We thank Yu. Litvinov for information concerning the GSI experiment.

%\newpage

\end{document}